\documentclass[twoside,11pt,leqno]{article}
\usepackage[english]{babel}
\RequirePackage{amsthm,multirow,amsmath,fancybox,amssymb}
\RequirePackage[dvips]{graphicx}

\newcommand{\R}{{\mathbb R}}

\newcommand{\RR}{{\mathbb R^2}}
\newcommand{\RRdR}{{\mathbb R ^2 \times \mathbb R}}
\newcommand{\x}{{\rm{\bf{x}}}}
\newcommand{\BB}{{\mathcal B}}
\newcommand{\1}{{\mathbf 1}}
\newtheorem{lemma}{Lemma}
\newtheorem{definition}{Definition}
\newtheorem{algorithm}{Algorithm}

\begin{document}

\thispagestyle{empty}
\begin{center}
{\Large\sc 
A multi-scale area-interaction model for spatio-temporal point patterns
}\\[.5in] 

\noindent
\large{A. Iftimi$^{1}$, M.N.M. van Lieshout$^{2,3}$, F. Montes$^1$}\\[0.2in]
\textit{$^1$Department of Statistics and Operations Research,
University of Valencia \\
C/ Doctor Moliner, 50, 46100 Burjassot-Valencia, Spain}\\
\textit{$^2$CWI, P.O. Box 94079, NL-1090 GB Amsterdam, The Netherlands}\\
\textit{$^3$Department of Applied Mathematics, University of Twente \\
P.O. Box 217, NL-7500 AE Enschede, The Netherlands}\\[0.5in]
\end{center}

\begin{verse}
{\footnotesize
\noindent
{\bf Abstract}\\
\noindent
Models for fitting spatio-temporal point processes should incorporate 
spatio-temporal inhomogeneity and allow for different types of interaction 
between points (clustering or regularity). This paper proposes an extension 
of the spatial multi-scale area-interaction model to a spatio-temporal 
framework. This model allows for interaction between points at different 
spatio-temporal scales and the inclusion of covariates. We fit the proposed 
model to varicella cases registered during 2013 in Valencia, Spain. The 
fitted model indicates small scale clustering and regularity for higher 
spatio-temporal scales.
\\[0.2in]

\noindent
{\em Keywords \& Phrases:}
Gibbs point processes; multi-scale area-interaction model; 
spatio-temporal point processes; varicella.

\noindent
{\em 2010 Mathematics Subject Classification:}
60D05, 60G55, 62M30. 
% point processes 60G55
% stochastic geometry 60D05
% spatial processes 62M30
}
\end{verse}

\section{Introduction}
\label{introduction}

Spatio-temporal patterns are increasingly observed in many different fields, 
including ecology, epidemiology, seismology, astronomy and forestry. The 
common feature is that all observed events have two basic characteristics: 
the location and the time of the event. Here we are mainly concerned with 
epidemiology \cite{epidemiology}, which studies the distribution, causes 
and control of diseases in a defined human population. The locations of 
the occurrence of cases give information on the spatial behavior of the 
disease, whereas the times, measured on different scales (days, weeks, 
years, period of times), give insights on the temporal response of the 
overall process. An essential point to take into consideration is that 
people are not uniformly distributed in space, hence information on the 
spatial distribution of the population at risk is crucial when analyzing 
spatio-temporal patterns of diseases.  

Realistic models to fit epidemiological data should incorporate 
spatio-temporal inhomogeneity and allow for different types of dependence 
between points. One important class of such models is the family of Gibbs 
point processes, defined in terms of their probability density function 
\cite{marie-colette,Ripley88,Ripley89}, and, in particular, the sub-class of
pairwise interaction processes. Well-known examples of pairwise interaction
processes are the Strauss model \cite{Kelly,Strauss} or the hard core process, 
a particular case of the Strauss model where no points ever come closer to 
each other than a given threshold. However, pairwise interaction models 
are not always a suitable choice for fitting clustered patterns. A family 
of Markov point processes that can fit both clustered and ordered patterns 
is that of the area- or quermass-interaction models 
\cite{lieshout_baddeley,kendall99}. These models are defined in terms
of stochastic geometric functionals and display interactions of all orders.
Methods for inference and perfect simulation are available in 
\cite{Dereudre14,Haggstrom99,Kendall98,moller10}.

Most natural processes exhibit interaction at multiple scales. The classical 
Gibbs processes model spatial interaction at a single scale, nevertheless 
\emph{multi-scale} generalizations have been proposed in the literature 
\cite{ambler10,gregori03,Picard09}. In this paper we propose an extension 
of the spatial multi-scale area-interaction model to a spatio-temporal 
framework. 

The outline of the paper is as follows. Section~\ref{section:preliminaries} 
provides some preliminaries in relation to notation and terminology. 
Section~\ref{section:space-time area int} gives the definition and Markov 
properties of our {spatio-temporal multi-scale area-interaction model}. 
Section~\ref{section:simulation} adapts simulation algorithms, such as the 
Metropolis-Hastings and the birth-and-death algorithms, to our context. 
Section~\ref{section:inference} treats the pseudo-likelihood method for 
inference as well as an extension of the Berman-Turner procedure. The ideas 
are illustrated on a simulated example. The model is applied to a varicella 
data set in Section~\ref{section:data}. Section~\ref{section:discussion} 
presents final remarks and a discussion of future work.

\section{Preliminaries}
\label{section:preliminaries}

A realization of a spatio-temporal point process consists of a finite number 
$n \geq 0$ of distinct points $(x_i,t_i)$, $i=1, \ldots, n$, that are observed 
within a compact spatial domain $W_S \subseteq \RR$ and time interval 
$W_T \subseteq \R$. The pattern formed by the points will be denoted by 
$\x = \{ (x_i,t_i) \}_{i=1}^n$. For a mathematically rigorous account, 
the reader is referred to \cite{daley03,daley08}.

We define the Euclidean norm $||x|| = ( x_1^2 + x_2^2 )^{1/2}$ and the 
Euclidean metric $d_{\RR}(x,y) = || x-y ||$ for $x = (x_1,x_2) \in \RR$ 
and $y = (y_1,y_2) \in\RR$. We need to treat space and time differently, 
thus on $\RRdR$ we consider the supremum norm 
$|| (x,t) ||_\infty = \max\{ ||x||, |t| \}$ and the supremum metric 
$d((x,t),(y,s)) = ||(x,t)-(y,s)||_\infty = \max\{ ||x-y||, |t-s| \}$, 
where $(x,t), (y,s)\in \RRdR$. Note that $(\RRdR,d(\cdot,\cdot))$ as well
as its restriction to $W_S \times W_T$ is a complete, separable metric space. 
We write $\BB(\RRdR) = \BB(\RR)\otimes \BB(\R)$ for the Borel 
$\sigma$-algebra and $\ell$ for Lebesgue measure. We denote by 
$\oplus$ the Minkowski addition of two sets $A, B\subset \RR$, 
defined as the set $A\oplus B = \{ a + b: a \in A, b\in B \}$.

As stated in Section~\ref{introduction}, Gibbs models form an important 
class of models able to fit epidemiological data exhibiting spatio-temporal 
inhomogeneity and interaction between points. In space, the Widom-Rowlinson 
\emph{penetrable sphere model} \cite{Widom} produces clustered point 
patterns; the more general area-interaction model \cite{lieshout_baddeley} 
fits both clustered and inhibitory point patterns. In its most simple form,
the area-interaction model is defined by its probability density
\begin{align}
\label{eq:spatial_area_int}
p(\x) = \alpha \lambda^{n(\x)} \gamma^{-A(\x)}
\end{align}
with respect to a unit rate Poisson process on $W_S$. Here $\alpha$ is the 
normalizing constant, $\x$ is a spatial point configuration in $W_S\subset\RR$,
$n(\x)$ is the cardinality of $\x$ and $A(\x)$ is the area of the union of 
discs of radius $r$ centered at $x_i\in \x$ restricted to $W_S$. The positive
scalars $\lambda$, $\gamma$ and $r>0$ are the parameters of the model. 
Note that, as emphasized in \cite{marie-colette}, Gibbsian interaction 
terms can be combined to yield more complex models. Doing so, 
\cite{ambler10,gregori03,Picard09} develop an extension of the 
area-interaction process which incorporates both inhibition and attraction. 
We propose a further generalization of the area-interaction model to allow 
multi-scale interaction in a spatio-temporal framework.

\section{Space-time area-interaction processes}
\label{section:space-time area int}

Let $\x$ be a finite spatio-temporal point configuration on $W_S\times 
W_T\subset\RRdR$, that is, a finite set of points, including the empty set.

\begin{definition} 
The \emph{spatio-temporal multi-scale area-interaction process} is the 
point process with density	
\begin{align}
\label{def_area_int}
p(\x) = \alpha \prod_{ (x,t) \in \x } \lambda (x,t) 
               \prod_{j=1}^{m} \gamma_j^{-\ell(\x\oplus G_j)}
\end{align}
with respect to a unit rate Poisson process on $W_S\times W_T$, where 
$\alpha>0$ is a normalizing constant, $\lambda\geq 0$ is a measurable and 
bounded function, $\ell$ is Lebesgue measure restricted to $W_S \times W_T$, 
$\gamma_j>0$ are the interaction parameters, $G_j$ are some compact 
subsets of\/ $\RRdR$ with size depending on $j$, $j = 1, \ldots, m$, 
$m\in \mathbb{N}$, and $\oplus$ denotes Minkowski addition. 
\end{definition}

Note that when $\x$ is the empty set, $p(\x) = \alpha$. The interaction 
parameters have the same interpretation as for the spatial area-interaction 
model \eqref{eq:spatial_area_int}. For fixed $j \in \{ 1, \ldots, m \}$, when 
$0 < \gamma_j < 1$ we would expect to see \emph{inhibition} between points at 
spatio-temporal scales determined by the definition of the compact set $G_j$. 
On the other hand, when $\gamma_j > 1$ we expect \emph{clustering} between the 
points. We observe that \eqref{def_area_int} reduces to an inhomogeneous 
Poisson process when $\gamma_j = 1$ for all $j \in \{ 1, \ldots, m \}$. 

Covariates can be introduced in the model by letting the intensity 
function $\lambda$ be a measurable and bounded function 
$\lambda(x,t) = \rho ( Z(x,t) )$ of the covariate vector $Z(x,t)$. 

The new model proposed in \eqref{def_area_int} successfully extends the 
area-interaction model to multi-scale interaction for spatio-temporal 
point patterns.

\begin{lemma}
The density \eqref{def_area_int} is measurable and integrable for all 
 $\gamma_j$, $j = 1, \ldots, m$, $m \in \mathbb{N}$.
\end{lemma}

\begin{proof}
Consider a point configuration, $\x$. Since $\ell$ is $\sigma$-finite and 
$G_j$ is compact, the map $\x \mapsto \ell(\x\oplus G_j)$ is measurable 
for any $j = 1, \ldots, m$. It follows that the map 
$\x \mapsto \exp[ - \ell(\x \oplus G_j) \log{\gamma_j} ]$ 
is measurable for any $j = 1, \ldots, m$. The map 
$\x \mapsto \prod_{x_i\in\x} \lambda(x_i,t_i)$ 
is also measurable by assumption, hence the density \eqref{def_area_int} is 
measurable.

To determine if \eqref{def_area_int} is integrable, we observe that 
$0 \leq \ell(\x\oplus G_j) \leq \ell(W_S \times W_T) < \infty$. The function 
$\lambda$ is integrable by assumption, hence \eqref{def_area_int} is 
dominated by an integrable function, and therefore integrable.
\end{proof}

As a further simplification, for fixed $j \in \{ 1, \ldots, m \}$, consider 
the case where 
$\x \oplus G_j = \bigcup_{ (x,t) \in \x } \mathcal{C}_{r_j}^{t_j}(x,t)$ 
is the union of all cylinders with radius $(r_j,t_j)$ centered in $(x,t)$ 
taken over all $(x,t)\in \x$. We define the cylinder with radius $(r_j,t_j)$ 
by
\begin{align*}
\mathcal{C}_{r_j}^{t_j}(x,t) =
  \{ (y,s) \in W_S \times W_T: ||x-y|| \leq  r_j, |t-s| \leq t_j \}.
\end{align*}
Then $\x\oplus G_j$ is the set of all points within the cylinders 
$\mathcal{C}_{r_j}^{t_j}(x,t)$ centered in points of $\x$ and the expression 
\eqref{def_area_int} reads
\begin{align}\label{model_with_r_and_t}
p(\x) = \alpha \prod_{ (x,t) \in \x } \lambda(x,t) 
     \prod_{j=1}^{m} \gamma_j^{ - \ell( \cup_{ (x,t) \in \x } \mathcal{C}_{r_j}^{t_j}(x,t) ) },
\end{align}
where $(r_j,t_j)$ are pairs of \emph{irregular parameters} 
\cite{baddeley_book} of the model and $\gamma_j$ are interaction parameters,
$j = 1, \ldots, m$. The function $\lambda$ is here assumed known for 
simplicity, but could also depend on further parameters.

\begin{figure}[htb]
\centering
\includegraphics[width=0.5\textwidth]{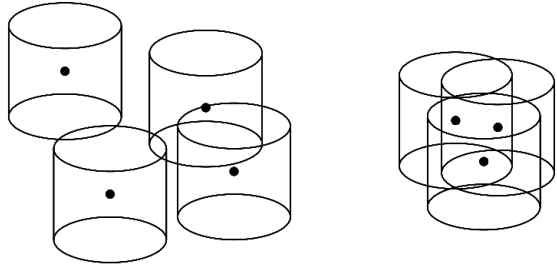}
\caption{An illustration of possible $\x\oplus G$ (cylinders around 
the points), where the black dots represent points of the process.}
\label{gamma}
\end{figure}  

Figure \ref{gamma} shows an illustration of $\x \oplus G_j$. 
When $0 < \gamma_j < 1$, point configurations such as the one on the 
left are likely to be observed (inhibition between points), whereas for 
large $\gamma_j>1$, point configurations such as the one on the right are 
more likely to be observed (attraction between points). 

\subsection{Markov properties}
\label{markov_properties}

Let $\sim$ on $\RRdR$ be a symmetric and reflexive relation on $\RRdR$, 
i.e.\ for any $(x,t), (y,s) \in \RRdR$, 
$(x,t) \sim (y,s) \Leftrightarrow (y,s) \sim (x,t)$ and 
$(x,t) \sim (x,t)$. Two points $(x,t)$ and $(y,s)$ are said to be 
\emph{neighbors} if $(x,t) \sim (y,s)$. An example of a fixed range 
relation on $\RRdR$ is 
\begin{align}\label{neighbour}
(x,t) \sim (y,s) \Leftrightarrow (x,t) \oplus G \cap (y,s) \oplus G 
  \neq \emptyset,
\end{align}
where $G = \mathcal{C}_{r_1}^{t_1}$ is a cylinder of radius $(r_1,t_1)$. 

\begin{definition}\label{def:markov_pp}
A point process has the \emph{Markov property} \cite{marie-colette,ripley_77}
with respect to the symmetric, reflexive relation $\sim$, if, for all point 
configurations $\x$ with $p(\x) > 0$, the following conditions are fulfilled:
\begin{enumerate}
\item $p({\rm{\bf{y}}}) > 0$ for all ${\rm{\bf{y}}} \subseteq \x$;
\item the likelihood ratio $\dfrac{ p(\x \cup \{(y,s)\}) }{ p(\x) }$
for adding a new point $(y,s)$ to a point configuration $\x$ depends 
only on points $(x,t)\in \x$ such that $(y,s) \sim (x,t)$, i.e.\ depends 
only on the neighbors of $(y,s)$. 
\end{enumerate} 
\end{definition}

\begin{lemma}
\label{lemma_markov}
The spatio-temporal multi-scale area-interaction process 
\eqref{def_area_int} is a Markov point process with respect to the 
relation \eqref{neighbour} in the sense of \cite{ripley_77}.
\end{lemma}

\begin{proof}
Note that if $p(\x) > 0$, since $\lambda(x,t) > 0$ for all $(x,t) \in \x$, 
then whenever ${\rm{\bf{y}}} \subseteq \x$, also $p({\rm{\bf{y}}}) > 0$. 
The likelihood ratio 
\begin{align}\nonumber
\dfrac{ p(\x \cup \{ (y,s) \} ) }{ p(\x) } &= 
\dfrac{ \alpha \displaystyle\left(\prod_{(x,t)\in \x} \lambda(x,t)\right)  
        \lambda(y,s)\displaystyle\prod_{j=1}^{m} 
           \gamma_j^{ - \ell( (\x \cup \{ (y,s) \} ) \oplus G_j ) 
}}{
   \alpha \displaystyle\prod_{(x,t)\in \x} \lambda(x,t) 
   \displaystyle\prod_{j=1}^{m} \gamma_j^{ - \ell( \x\oplus\it G_j ) } } \\ 
 &= \lambda(y,s) \displaystyle\prod_{j=1}^{m} 
    \displaystyle\gamma_j^{ - \ell\left( ((y,s)\oplus G_j)\setminus (\x\oplus G_j) \right)}.
\label{markov}
\end{align}
	
Note that 
\begin{align*}
( (y,s) \oplus G_j ) \setminus ( \x \oplus G_j) &=
( (y,s) \oplus G_j ) \cap 
   \left[ \bigcup_{(x,t)\in \x} (x,t) \oplus G_j \right]^C \\
&= ( (y,s)\oplus G_j) \cap 
   \left[ \bigcup_{(x,t) \sim (y,s)} (x,t) \oplus G_j \right]^C, \quad
\forall j = 1, \ldots, m.
\end{align*}
Thus \eqref{markov} depends only on the newly added point $(y,s)$ and its 
neighbors. Hence \eqref{def_area_int} defines a Markov point process with 
respect to $\sim$. 
\end{proof}

It follows that the density $p(\cdot)$ in \eqref{def_area_int} is Markov at 
range $2 \max \{ (r_j,t_j) \}$, $j=1, \ldots, m$.

Define the Papangelou conditional intensity of a point process with density 
$p$ by
\begin{align*}
\lambda((y,s); \x) = \dfrac{ p(\x \cup \{(y,s)\}) }{ p(\x) },
\end{align*}
whenever $p(\x)>0$ and $(y,s) \in \x$. Then, for the spatio-temporal 
multi-scale area-interaction process, by the proof of Lemma~\ref{lemma_markov}
we obtain that 
\begin{align}
\lambda((y,s); \x) &= \lambda(y,s) \displaystyle\prod_{j=1}^{m}  
\gamma_j^{-\ell(\mathcal{C}_{r_j}^{t_j}(y,s)
   \setminus \bigcup_{(x,t)\in \x} \mathcal{C}_{r_j}^{t_j}(x,t))},
\label{conditional_int1}
\end{align}
or, upon transformation to a logarithmic scale,
\begin{align*}
\log{\lambda((y,s); \x)} &= \log{\lambda(y,s)} - \displaystyle\sum_{j=1}^{m}  
(\log{\gamma_j}) \, \ell\left( \mathcal{C}_{r_j}^{t_j}(y,s) 
\setminus \bigcup_{(x,t)\in \x} \mathcal{C}_{r_j}^{t_j}(x,t) \right).
\end{align*}
Note that $\lambda(y,s)$ may be $0$, thus making $\log{\lambda(y,s)}$ 
ill-defined.

Write $\eta_j = \log{\gamma_j}$. Then, whenever well-defined,
\begin{align}\nonumber
\log{\lambda((y,s);\x)} &= 
\log{\lambda(y,s)} - \displaystyle\sum_{j=1}^{m}  \eta_j 
\int_{ W_S \times W_T } \1\{ (z,u) \in \mathcal{C}_{r_j}^{t_j}(y,s) \setminus 
  \bigcup_{(x,t)\in \x} \mathcal{C}_{r_j}^{t_j}(x,t) \} \, dz \, du\\
&= \log{\lambda(y,s)} - \displaystyle\sum_{j=1}^{m}  
  \int_{ \mathrm{F}_{r_j}^{t_j}(y,s) } \sum_{i=j}^{m} \eta_i \,
     \1\{ (z,u) \notin \bigcup_{(x,t) \in \x} \mathcal{C}_{r_i}^{t_i}(x,t) \} 
  \, dz \, du, 
\label{conditional_int2}
\end{align}
where $\mathrm{F}_{r_j}^{t_j}\it(x,t)$ is the difference between two concentric 
cylinders $\mathcal{C}_{r_j}^{t_j}(x,t)$ and $\mathcal{C}_{r_{j-1}}^{t_{j-1}}(x,t)$.

\begin{figure}[hbt]
\centering
\includegraphics[width=0.4\textwidth]{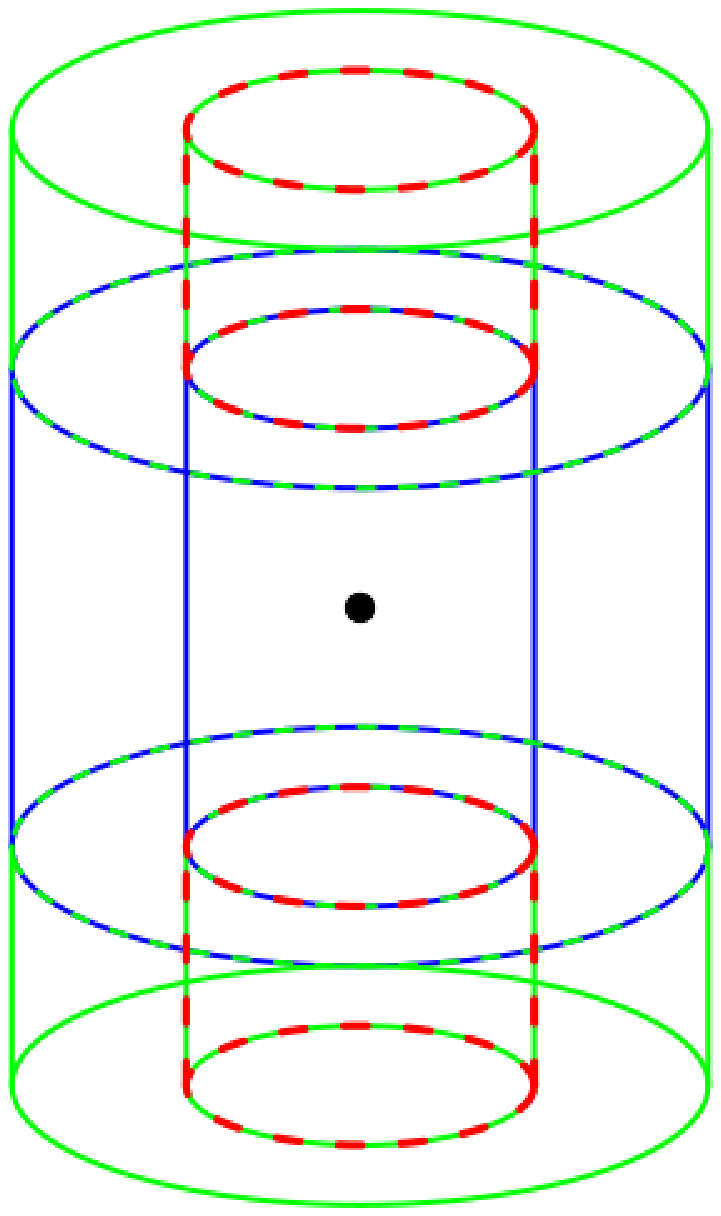}
\includegraphics[width=0.5\textwidth]{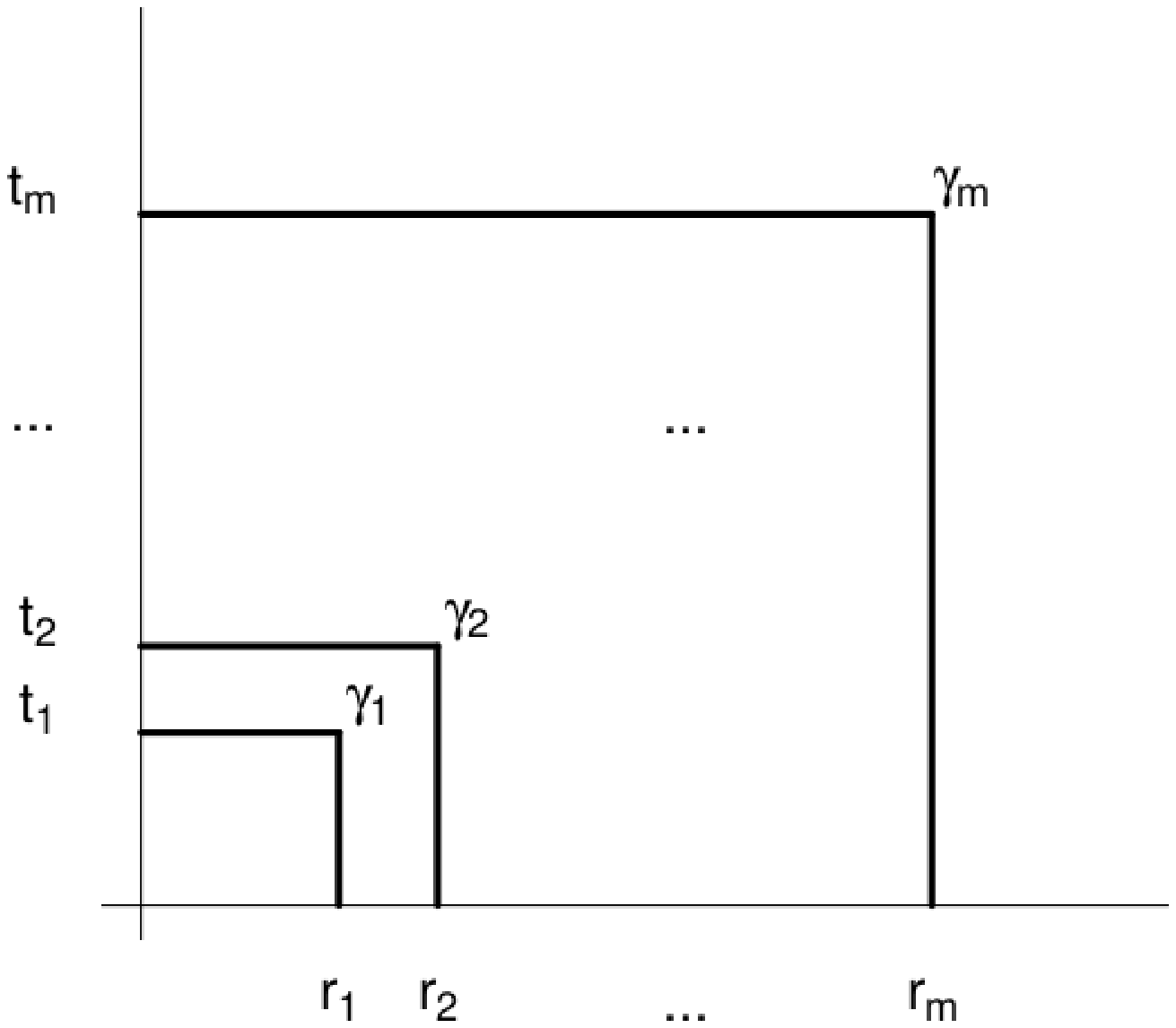}
\caption{(Left) An illustration of $F_{r_j}^{t_j}$ where the blue annulus 
corresponds to 
$\{ (y,s) \in W_S \times W_T: r_{j-1} < ||x-y|| \leq r_j, |t-s| \leq t_{j-1} \}$, 
the two green annuli represent 
$\{ (y,s) \in W_S \times W_T: r_{j-1} < ||x-y|| \leq r_j,
     t_{j-1} < |t-s| \leq t_{j} \}$ 
and the two red cylinders are  
$\{ (y,s) \in W_S \times W_T: ||x-y|| \leq r_{j-1}, t_{j-1} < |t-s|\leq t_{j} \}$.
(Right) Multi-scale behavior.}
\label{fig:F_cylinder}
\end{figure}

Indeed,
\begin{eqnarray*}
\mathrm{F}_{r_j}^{t_j}(x,t) & = & 
\mathcal{C}_{r_j}^{t_j}(x,t) \setminus \mathcal{C}_{r_{j-1}}^{t_{j-1}}(x,t) 
  \\
& = & \left\{ (y,s) \in W_S \times W_T: 
\begin{array}{l}
 r_{j-1} < ||x-y|| \leq r_j, |t-s| \leq t_{j-1} \ \mathrm{or} \\
 r_{j-1} < ||x-y|| \leq r_j, t_{j-1} < |t-s| \leq t_{j} \ \mathrm{or} \\
 ||x-y|| \leq r_{j-1}, t_{j-1} < |t-s| \leq t_{j} 
\end{array}
\right\}, 
\end{eqnarray*}
with $0 = r_0 < r_1 < \cdots < r_m$ and $0 = t_0 < t_1 < \cdots <t_m$. 
The left-most panel of Figure~\ref{fig:F_cylinder} shows an 
illustration of $F_{r_j}^{t_j}$ for fixed $r_j, t_j$. The blue annulus 
corresponds to 
$\{ (y,s) \in W_S \times W_T: r_{j-1} < ||x-y|| \leq r_j, |t-s| \leq t_{j-1} \}$,
the two green annuli represent 
$\{ (y,s) \in W_S \times W_T: r_{j-1} < ||x-y|| \leq r_j, 
  t_{j-1} < |t-s| \leq t_{j} \}$ 
and the two red cylinders form  
$\{ (y,s) \in W_S \times W_T: ||x-y|| \leq r_{j-1}, 
  t_{j-1} < |t-s| \leq t_{j} \}$. 
If, for $(y,s), ||y-x||>2 r_m, |s-t|>2 t_m, \forall (x,t)\in \x$, then 
\begin{align*}
\log{\lambda((y,s);\x)} &= \log{\lambda(y,s)} - 
\displaystyle\sum_{j=1}^{m} \left(\sum_{i=j}^{m} \eta_i\right) 
\ell(\mathrm{F}_{r_j}^{t_j}(y,s))\\
&= \log{\lambda(y,s)} - 
\displaystyle\sum_{j=1}^{m} \eta_j \ell(C_{r_j}^{t_j}(y,s)).
\end{align*}
To conclude this discussion, note that, in accordance with 
\cite{gregori03}, 
\begin{align*}
p(\x) = \alpha \prod_{(x,t)\in\x} \lambda(x,t) \,
 \exp{[ - \sum_{j=1}^{m} \alpha_j \, \ell(F_{r_j}^{t_j}(\x)) ] },
\end{align*}
where $\alpha_j = \sum_{i\geq j} \eta_i$  and 
$F_{r_j}^{t_j}(\x)) = (\x\oplus G_j) \backslash (\x\oplus G_{j-1})$ 
As before, $ G_j=C_{r_j}^{t_j}$.

The model in \eqref{def_area_int} with Papangelou conditional intensity 
defined by \eqref{conditional_int2} allows for models whose interaction
behavior varies across spatio-temporal scales, for example, inhibition at 
small scales, attraction at larger scales and randomness beyond.
The different spatio-temporal scales, $(r_j,t_j)$, are defined 
according to $\mathrm{F}_{r_j}^{t_j}$. Indeed, a point $(z,u)$ in 
$F_{r_j}^{t_j}(\x)$ contributes a term $\alpha_j$ to the energy
(the negative of the exponential term) in $p(\x)$. The right-most 
panel of Figure \ref{fig:F_cylinder} shows a visual representation 
of this multi-scale behavior. 

An important property of Markov densities is the fact that the Papangelou 
conditional intensity, $\lambda((y,s); \x)$, depends only on $(y,s)$ and its 
neighbors in $\x$, and is computationally convenient. This property will be 
exploited in the next section to design simulation algorithms for generating
realisations of the model.

\section{Simulation}
\label{section:simulation}

\subsection{The Metropolis-Hastings algorithm}
\label{subsection:mh_algorithm}

Consider a Markov point process on $W_S\times W_T\subseteq\RRdR$ defined by 
its density $p(\cdot)$. The \emph{Metropolis-Hastings algorithm}, first 
introduced in statistical physics \cite{barker65,metropolis}, is a tool for
constructing a Markov process with limit distribution defined by $p(\cdot)$.

Metropolis-Hastings algorithms are \emph{discrete time Markov} processes 
where transitions are defined as the proposal of a new state that is 
accepted or rejected based on the likelihood of the proposed state compared 
with the old state. We consider two types of proposals: addition (\emph{birth})
and deletion (\emph{death}) of a point. The likelihood ratio of the new 
state in comparison with the old state, for these type of transitions, is 
the (reciprocal) conditional intensity.

More precisely, consider the point configuration $\x$. We can 
propose either a birth or a death with respective probabilities $q(\x)$ 
and $1-q(\x)$ that depend on $\x$. For a birth, a new point $u \in W_S
\times W_T$ is sampled from a probability density $b(\x,\cdot)$ and the 
new point configuration $\x \cup \{ u \}$ is accepted with probability 
$A(\x, \x\cup \{ u \})$, otherwise the state remains unchanged, $\x$. 
For a death, the point $x\in\x$ chosen to be eliminated is selected 
according to a discrete probability distribution $d(\x,\cdot)$ on $\x$, 
and the proposal $\x\setminus \{ x \}$ is accepted with probability 
$A(\x, \x \setminus \{ x \})$, otherwise the state remains unchanged.

In general, we can choose $b(\cdot,\cdot), d(\cdot,\cdot)$ and $q(\cdot)$ 
as we prefer. However, an important condition to consider is that of 
\emph{detailed balance}, and therefore time-reversibility of the Markov 
process,
\begin{align}
\label{detailed_balance}
\nonumber
& q(\x) \, b(\x, u) \, A(\x, \x \cup \{ u \}) \, p(\x) = \\ 
& ( 1 - q(\x \cup \{ u \}) ) \, d(\x \cup \{ u \}, u) \,
  A(\x \cup \{ u \}, \x) \, p(\x \cup \{ u \}).
\end{align}
For simplicity, consider the case that births and deaths are equally 
likely and sampled uniformly, that is, $q \equiv 1/2$, 
$b \equiv 1 / \ell(W_S \times W_T)$ and $d({\x}, \cdot) = 1 / n(\x)$,
where $n(\x)$ is the number of points in the point configuration $\x$. 
Then \eqref{detailed_balance} reduces to 
\begin{align*}
\dfrac 1 2 \dfrac{1}{\ell(W_S \times W_T)} {A(\x, \x \cup \{ u \}) \, p(\x)} 
&= \left(1-\dfrac 1 2\right) \dfrac{1}{n({\x})+1}
 A(\x \cup \{ u \}, \x) \, p(\x \cup \{ u \}) \\
\dfrac{1}{\ell(W_S \times W_T)} {A(\x, \x \cup \{ u \}) \, p(\x)} &=  
\dfrac{1}{n({\x})+1} A(\x \cup \{ u \},\x) \, p(\x \cup \{ u \}) \\
\dfrac{A(\x,\ x \cup \{ u \})}{A(\x \cup \{ u \}, \x)} &= 
\underbrace{\dfrac{\ell(W_S \times W_T)}{n({\x})+1} 
\times \dfrac{p(\x \cup \{ u \})}{p(\x)}}_{=r(\x,u)} .
\end{align*}
Thus, more likely configurations can be favored by setting 
${A(\x, \x \cup \{ u \})} = \min\{ 1, r(\x, u) \}$, and  
${A(\x \cup \{ u \}, \x)} = \min\{ 1, 1/r(\x, u) \}$. Therefore,
using equation \eqref{conditional_int1}, for the spatio-temporal
multi-scale area-interaction process \eqref{def_area_int}, the 
ratio $r(\x, u)$ for $u = (y,s)$ reduces to
\begin{align}\label{metropolis}
r(\x,u) = \dfrac{\ell(W_S \times W_T)}{n({\x})+1} 
  \lambda(y,s) \displaystyle\prod_{j=1}^{m} 
  \gamma_j^{-\ell(\mathcal{C}_{r_j}^{t_j}(y,s) 
     \setminus \bigcup_{(x,t)\in \x} \mathcal{C}_{r_j}^{t_j}(x,t))}.
\end{align}

In practice, we will use the logarithmic form of the conditional 
intensity as given in equation \eqref{conditional_int2}. When the 
region $W_S\times W_T$ is irregular we use rejection sampling to 
generate a point uniformly at random from $W_S \times W_T$.

\subsection{Birth-and-death-processes}

In this section we discuss methods for simulating \eqref{def_area_int} using 
birth-and-death processes \cite{preston77}. The birth-and-death process is 
a \emph{continuous time Markov} process where the transition from one state 
to another is given by either a \emph{birth} or a \emph{death}. A birth is 
the transition from a point configuration 
$\x \in W_S \times W_T \subseteq \RRdR$ to $\x \cup \{ u \}$ by adding the 
point $u \in W_S \times W_T$. A death is the transition from a point 
configuration $\x$ to $\x \setminus \{x\}$ by eliminating a point $x \in \x$. 
We denote by $b(\x, u) \, du$ the transition rate for a birth and by 
$d(\x, x)$ the transition rate of a death. The total birth rate from $\x$
is the integral 
\[
B(\x) = \int_{W_S \times W_T} b(\x, u) \, du
\]
and the total death rate is
\[
D(\x) = \sum_{x\in \x} d(\x, x). 
\]
The process stays in state $X^{(n)} = \x$ for an exponentially distributed 
random sojourn time $T^{(n)}$ with mean $1/(B(\x)+D(\x))$. The detailed 
balance equations are given by 
\begin{align}\label{balance_eq}
b(\x, u) \, p(\x) = d(\x \cup \{ u \}, u) \, p(\x \cup \{ u \}).
\end{align}

We consider the particular case when the death rate is constant 
\cite{ripley_s77}, $d({\x}, x) = 1$. Hence, for the spatio-temporal
multi-scale area-interaction process (\ref{def_area_int}), the birth 
rate is given by the conditional intensity (cf.\ equation
\eqref{conditional_int1})
\begin{align}\label{birth_rate}
b(\x, (y,s)) = \dfrac{p(\x \cup \{ (y,s) \})}{p(\x)} =
\lambda(y,s) \displaystyle\prod_{j=1}^{m}  
   \gamma_j^{-\ell(\mathcal{C}_{r_j}^{t_j}(y,s) \setminus \bigcup_{(x,t)\in \x} 
                   \mathcal{C}_{r_j}^{t_j}(x,t))}.
\end{align}

For computation of the ratio in equation \eqref{birth_rate} we will use 
the logarithmic form of the conditional intensity as in equation 
\eqref{conditional_int2}.

Following \cite{marie-coletteSTG, marie-colette} we define an algorithm 
for simulating a birth-and-death process and generate the successive states 
$X^{(n)}$ and the sojourn times $T^{(n)}$ as detailed in 
Algorithm~\ref{algorithm_bd} which incorporates a rejection sampling step
for computational convenience. Define a threshold $w(\x)$, and,
for $u \notin \x$, set 
\[
g(\x, u) = 
\begin{cases}
b(\x, u), & \text{if } b(\x, u) \geq w(\x) \\
w(\x),    & \text{otherwise}.
\end{cases}
\]
A common choice is to take $w(\x)$ equal to an upper bound to the 
conditional intensity.

Denote by $G(\x)$ the integral of $g$. We generate the sequence of
$(X^{(n)},T^{(n)})$ as follows. 
\begin{algorithm}\label{algorithm_bd}
Initialize $X^{(0)} = \x_0$ for some finite point configuration with
density function $p(\x_0) > 0$. For $n = 0, 1, \ldots$, if $X^{(n)}=\x$, 
compute $D = D(\x)$, $G = G(\x)$ and set $T^{(n)} = 0$.
\begin{itemize}
\item Add an exponentially distributed time to $T^{(n)}$ with mean $1/(D+G)$;
\item with probability $D/(D+G)$ generate a death
$X^{(n+1)} = \x \setminus \{ x\}$ by eliminating one of the current points 
$x\in \x$ at random according to distribution $d(\cdot,\cdot)$ and stop;
\item else sample a point $u$ from $g(\x, u) / G$; with probability 
$b(\x, u) / g(\x, u)$ accept the birth $X^{(n+1)} = \x \cup \{ u\}$ and stop;
otherwise repeat the whole algorithm.
\end{itemize}
\end{algorithm}

\section{Inference}
\label{section:inference}

\subsection{Pseudo-likelihood method}
\label{subsection:pl}

In this section, we assume that the function $\lambda$ is known and denote 
by $\theta = (\gamma_1, \gamma_2, \ldots, \gamma_m)$ the interaction parameters 
in model \eqref{model_with_r_and_t}. To estimate $\theta$, we may use 
pseudo-likelihood which aims to optimize 
\begin{align}\label{pseudo_likelihood}
PL(\x, \theta) =   \exp\left(
    -\int_{W_S} \int_{W_T} \lambda_\theta((u,v); \x) \, du \, dv \right) 
\prod_{(x,t) \in \x} \lambda_\theta((x,t); \x \setminus \{ (x,t) \}),
\end{align}
where $\lambda_\theta((u,v); \x)$ is the conditional intensity that depends 
on $\theta$ \cite{besag75}. 

For a Poisson process the conditional intensity is equal to the intensity 
function, hence pseudo-likelihood is equivalent to maximum likelihood. In 
general, the pseudo-likelihood $PL(\x,\theta)$ is only an approximation of 
the true likelihood. However, no sampling is needed and the computational 
load will be considerably smaller than for the maximum likelihood method.

The maximum pseudo-likelihood normal equations are then given by
\begin{align}
\label{pl_eq}
\dfrac{\partial}{\partial\theta} \log PL(\x, \theta) = 0,
\end{align}
where 
\begin{align}\label{max_pl}
\log PL(\x, \theta) = \sum_{(x,t)\in \x} 
   \log \lambda_\theta ((x,t); \x \setminus \{ (x,t) \})
 - \int_{W_S} \int_{W_T} \lambda_\theta((u,v); \x) \, du \, dv.
\end{align}

As seen in Section~\ref{markov_properties}, the Papangelou conditional 
intensity of the spatio-temporal multi-scale area-interaction model is
\begin{align*}
\lambda_\theta((y,s); \x) = \lambda(y,s) \displaystyle\prod_{j=1}^{m}  
\gamma_j^{-\ell(\mathcal{C}_{r_j}^{t_j}(y,s) \setminus \cup_{\x}^j)},
\end{align*}
where $\cup_{\x}^j = \bigcup_{(x,t) \in \x} \mathcal{C}_{r_j}^{t_j}(x,t)$;
its logarithm reads
\begin{align*}
\log{\lambda_\theta((y,s);\x)} &= \log{\lambda(y,s)} - 
\displaystyle\sum_{j=1}^{m}  (\log{\gamma_j}) \, \ell(\mathcal{C}_{r_j}^{t_j}(y,s)
\setminus \cup_{\x}^j).
\end{align*}

Following \cite{baddeley_turner} we denote by $S_j(y,s)=
\ell(\mathcal{C}_{r_j}^{t_j}(y,s) \setminus \cup_{\x}^j)$ the sufficient 
statistics, hence 
$\log{\lambda_\theta((y,s); \x)} = \log \lambda(y,s) - \theta^T \begin{bmatrix}
S_1(y,s) \\ \cdots \\ S_m(y,s)
\end{bmatrix}$. This notation will be further used in 
Algorithm~\ref{algorithm_mpl}. 

Thus, equation \eqref{pl_eq} gives us the pseudo-likelihood equations
\begin{align}\label{eq_log_pl}
\nonumber
\dfrac{\partial}{\partial\theta} \big( \sum_{(x,t) \in \x}  
& \left[ \log{\lambda(x,t)} - \displaystyle\sum_{j=1}^{m} (\log{\gamma_j}) \,
    \ell(\mathcal{C}_{r_j}^{t_j}(x,t) \setminus \cup_{\x \setminus \{ (x,t) \}}^j) 
  \right] - \\
&   - \int_{W_S} \int_{W_T} \lambda(u,v) \displaystyle\prod_{j=1}^{m}  
\gamma_j^{ - \ell(\mathcal{C}_{r_j}^{t_j}(u,v) \setminus \cup_{\x}^j)} \, du \, dv \big)
= 0. 
\end{align}

For every parameter $\gamma_i$, $i = 1, 2, \ldots, m$, the equations 
\eqref{eq_log_pl} read
\begin{align}\label{final_eq_pl}
\sum_{(x,t) \in \x} \dfrac{\ell(\mathcal{C}_{r_i}^{t_i}(x,t) \setminus 
  \cup_{\x \setminus \{ (x,t) \}}^i))}{\gamma_i} = 
\int_{W_S} \int_{W_T}  \lambda(u,v)
  \dfrac{\ell(\mathcal{C}_{r_i}^{t_i}(u,v) \setminus \cup_{\x}^i)}{\gamma_i} 
\prod_{j=1}^m \gamma_j^{ - \ell(\mathcal{C}_{r_j}^{t_j}(u,v) \setminus \cup_{\x}^j)}
\, du \, dv.
\end{align}

The major difficulty is to estimate the integrals on the right hand side of
equations \eqref{final_eq_pl}. Baddeley and Turner \cite{baddeley_turner} 
propose using the Berman-Turner method to approximate the integral in 
\eqref{max_pl} by
\begin{align*}
\int_{W_S} \int_{W_T} \lambda_\theta ((u,v);\x) \, du \, dv \approx 
\sum_{j=1}^{n} \lambda_\theta((u_j,v_j); \x) \, w_j, 
\end{align*} 
where $(u_j,v_j)$ are points in $W_S \times W_T$ and $w_j$ are quadrature 
weights. This yields an approximation for the log pseudo-likelihood of the 
form
\begin{align}\label{eq_log_pl_1}
\log PL(\x,\theta) \approx \sum_{(x,t)\in \x} 
  \log \lambda_\theta ((x,t); \x \setminus \{ (x,t) \}) -
\sum_{j=1}^{n} \lambda_\theta((u_j,v_j); \x) \, w_j.
\end{align} 

Note that if the set of points $\{ (u_j,v_j), j = 1, \ldots, n \}$ includes all 
the points $(x,t) \in \x$, we can rewrite \eqref{eq_log_pl_1} as 
\begin{align}\label{final_eq_log_pl}
\log PL(\x,\theta) \approx \sum_{j=1}^{n} (y_j \log \lambda_j - \lambda_j) \, w_j,
\end{align}
where $\lambda_j = \lambda_\theta((u_j,v_j); \x \setminus \{ (u_j, v_j) \})$, 
$y_j = z_j / w_j$ and 
\begin{align}\label{indicators}
z_j = 
\begin{cases}
1,  & \text{if } (u_j,v_j) \in \x \ (\mathrm{ is\  a\ point}),\\
0,  & \text{if } (u_j,v_j) \notin \x \ (\mathrm{ is\  a\ dummy \ point}).
\end{cases}
\end{align}
The right hand side of \eqref{final_eq_log_pl}, for fixed $\x$, is formally 
equivalent to the log-likelihood of independent Poisson variables 
$Y_j\sim \mathrm{Poisson}(\lambda_j)$ taken with weights $w_j$. Therefore 
\eqref{final_eq_log_pl} can be maximized using software for fitting 
generalized linear models. 

In summary, the method is as follows.

\begin{algorithm}\label{algorithm_mpl} 
\begin{itemize}
\item Generate a set of dummy points and merge them with all the data 
points in $\x$ to construct the set of quadrature points $(u_j,v_j) \in 
W_S \times W_T$;
\item compute the quadrature weights $w_j$;
\item obtain the indicators $z_j$ defined in \eqref{indicators} and 
calculate $y_j = z_j / w_j$;
\item compute the values $S_j(u_j,v_j)$ of the sufficient statistics at 
each quadrature point;
\item fit a generalized log-linear Poisson regression model with
parameters $\log \lambda_j$ given by
$\log\lambda(u_j, v_j) - \theta^T S( u_j, v_j)$, responses $y_j$ and 
weights $w_j$.
\end{itemize}
\end{algorithm}

The coefficient estimates returned by Algorithm \ref{algorithm_mpl} give 
the maximum pseudo-likelihood estimator $\hat{\theta}$ for $\theta$.

In order to estimate the parameters 
$\theta = (\gamma_1, \gamma_2, \ldots, \gamma_m)$ using the above method 
we need to have values for the \emph{irregular} parameters $r_j$ and 
$t_j$ for $j = 1, \ldots, m$. Baddeley and Turner \cite{baddeley_turner} 
suggest fitting the model for a range of values of these parameters and 
choose the values which maximize the pseudo-likelihood. Additionally,
we recommend to first compute some summary statistics, such as the pair 
correlation or auto-correlation function, to narrow down the search. 

We construct the quadrature scheme as a partition of $W_S \times W_T$ 
dividing the spatio-temporal area into cubes $C_k$ of equal volume. In 
the center of each cube $C_k$ we place exactly one dummy point. We then 
assign to each dummy or data point $(u_j,v_j)$ a weight $w_j= v /n_j$ where 
$v$ is the volume of each cube, and $n_j$ is the number of points, dummy 
or data, in the same cube as $(u_j,v_j)$. These weights are called the 
counting weights \cite{baddeley_turner}.  

We conclude this section by mentioning briefly an alternative way to define 
the quadrature scheme (Algorithm~\ref{algorithm_mpl}). Indeed,
\cite{baddeley_turner} suggest the use of a Dirichlet tessellation to
generate the quadrature weights. A quadrature scheme generated this way 
would mean that the weight of each point would be equal to the volume of the 
corresponding Dirichlet 3-dimensional cell. The computational cost of such 
a method is very high. Therefore, in this paper, we partition $W_S \times W_T$ 
into cubes of equal volume, as described above. 

\subsection{Simulation and parameter estimation of a 
spatio-temporal area interaction process}

For illustration purposes, we simulate two multi-scale spatio-temporal 
area interaction processes as defined in  \eqref{model_with_r_and_t}, 
one which exhibits small scale inhibition and large scale clustering 
(\emph{simulation 1}) and a second one which exhibits small scale 
clustering and large scale inhibition (\emph{simulation 2}).

We consider the spatio-temporal domain 
$W_S \times W_T = ([0,1] \times [0,1]) \times[0,1]$ and in both cases take 
constant $\lambda\equiv 50$. For the irregular parameters we choose the 
same spatio-temporal scales $r_1 = 0.03$, $r_2 = 0.05$, $t_1 = 0.03$ and 
$t_2 = 0.05$ for both simulations. We use the Metropolis-Hastings 
algorithm described in Section \ref{subsection:mh_algorithm} with $20,000$ 
iterations implemented in the \texttt{MPPLIB C++} library \cite{mppblib}. To 
estimate the parameters we follow the steps in Algorithm~\ref{algorithm_mpl}. 
We partition $W_S\times W_T$ into $10^3 = 1,000$ cubes of volume $10^{-3}$. 
In the center of each cube we place a dummy point, obtaining a total of 
$1,000$ dummy points. We then compute the sufficient statistics 
for each data and dummy point using the \texttt{MPPLIB C++} library and 
apply Algorithm~\ref{algorithm_mpl} to obtain the estimates for the parameters. 
For the implementation of the pseudo-likelihood method we use the statistical 
software \texttt{R} \cite{R_software} together with the \texttt{spatstat} 
\cite{baddeley_book} package. The theoretical background for computing the
`envelopes', that is the confidence interval bounds given as $2.5\%$ and 
$97.5\%$ in Tables \ref{tab:param_sim1} and \ref{tab:param_sim2} for a 
Poisson process is exhaustively described in \cite{kutoyants}.

\begin{figure}[htb]
\centering
\includegraphics[width=0.45\textwidth]{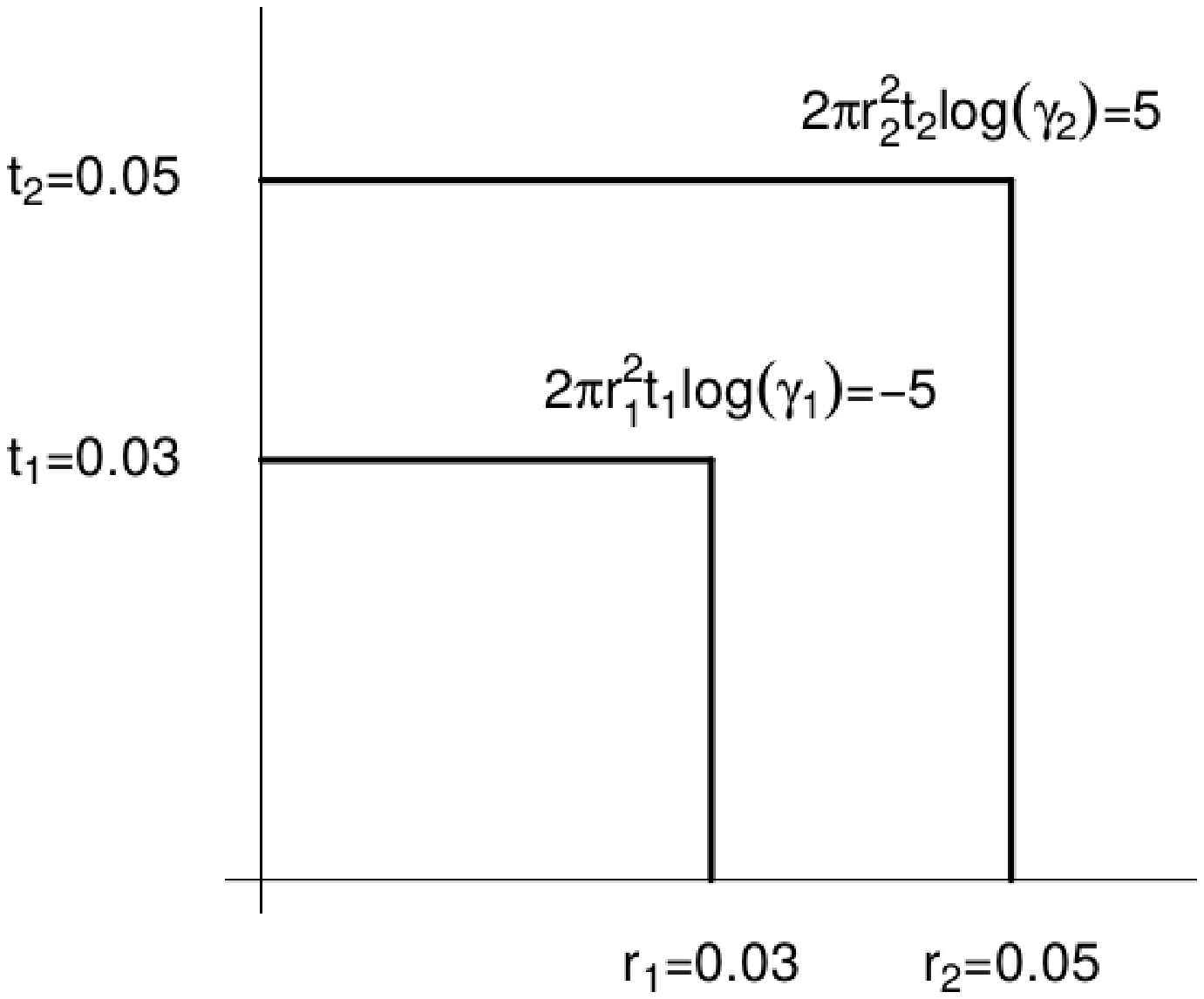}
\includegraphics[width=0.45\textwidth]{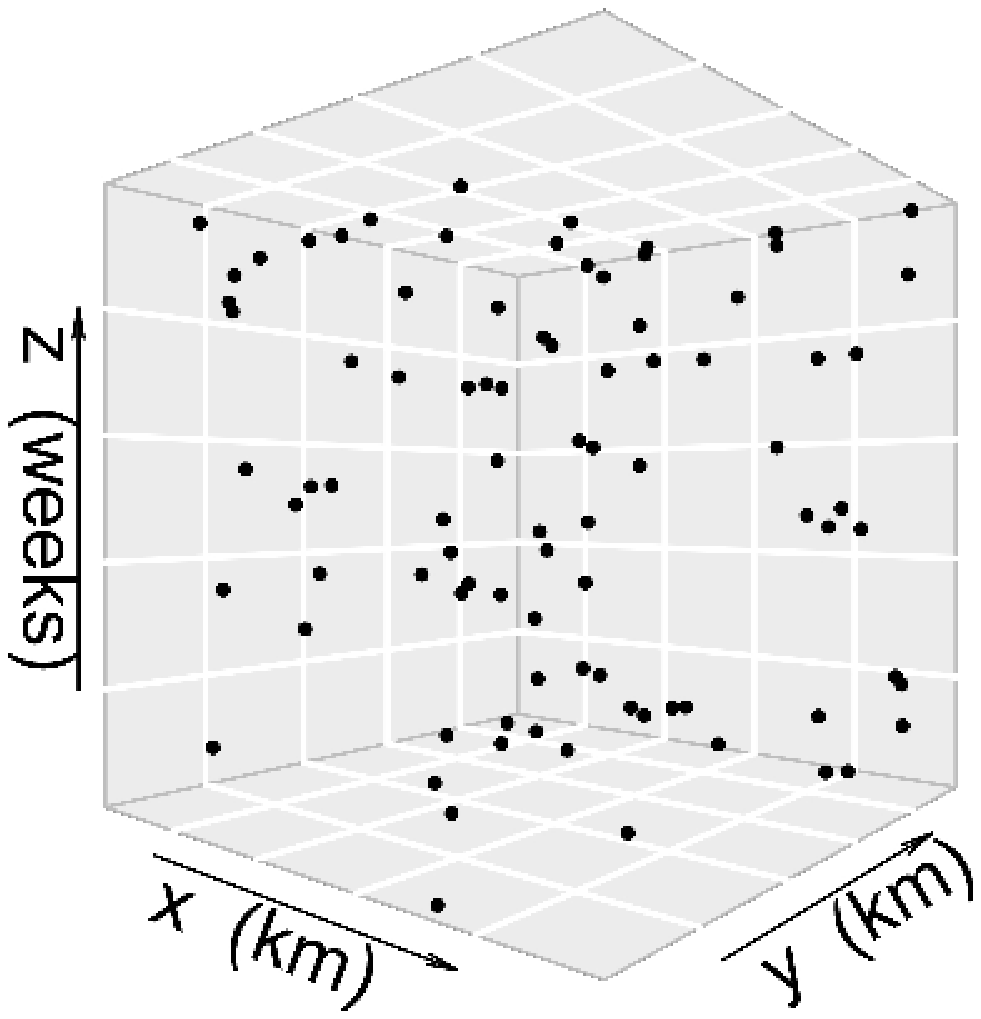}
\includegraphics[width=0.45\textwidth]{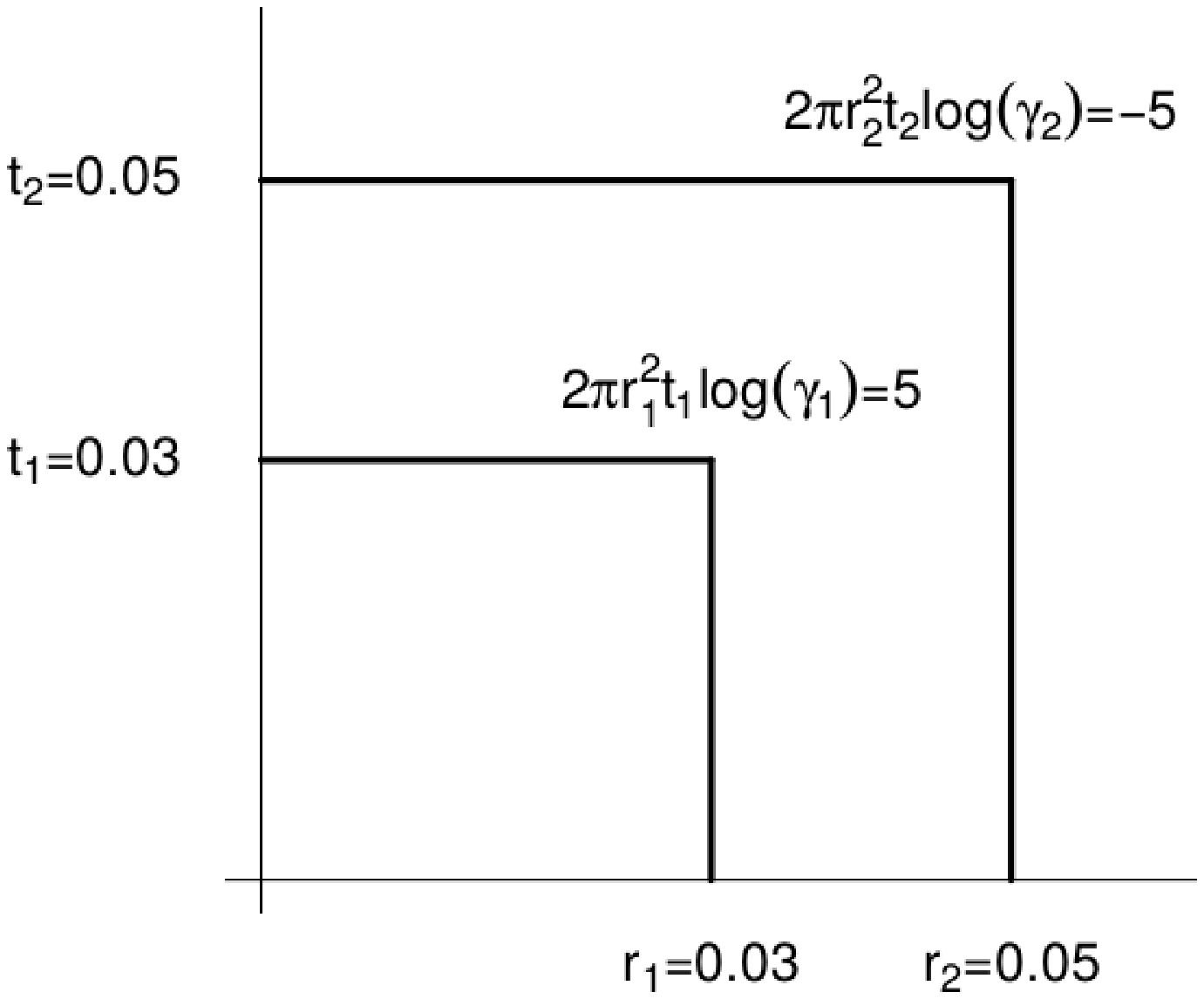}
\includegraphics[width=0.45\textwidth]{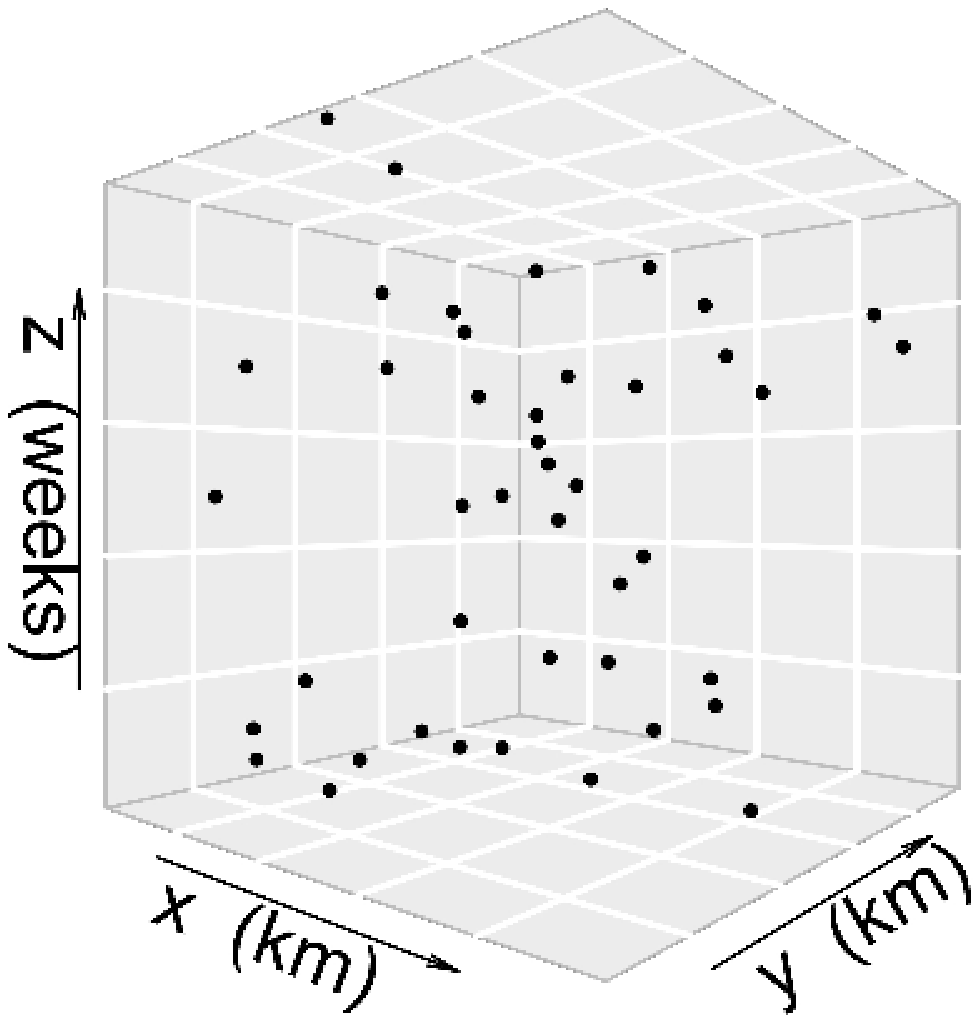}
\caption{(Top left) Model parameters for \emph{simulation 1}. 
(Top right) A realization of the first model. 
(Bottom left) Model parameters for \emph{simulation 2}. 
(Bottom right) A realization of the second model.}
\label{fig:gamma_sim1-2}
\end{figure}

Figure \ref{fig:gamma_sim1-2} (top left) shows the interaction parameters 
for \emph{simulation 1}, $2\pi r_1^2t_1\log(\gamma_1)=-5$ and 
$2\pi r_2^2t_2\log(\gamma_2)=5$. This setting of parameters gives us the 
spatio-temporal point configuration shown in the top right panel of
Figure \ref{fig:gamma_sim1-2} which indeed shows small scale inhibition between 
points and large scale clustering. The parameters estimates are shown in 
Table \ref{tab:param_sim1}.

\clearpage 

\begin{table}[hbt]
\begin{center}
\begin{tabular}{|r|rrr|} 
\hline
                                & Estimate   & 2.5 \% &  97.5 \% \\
\hline
$\log\lambda$	                &  6.07 &   3.57 & 8.02 \\ 
$2\pi r_1^2t_1\log(\gamma_1)=-5$ & -2.45 &	-5.48 & 0.37 \\
$2\pi r_2^2t_2\log(\gamma_2)=5$  &  4.48 &   2.44 & 6.48 \\
\hline
\end{tabular}
\end{center}
\caption{Parameter estimates for simulation 1.}
\label{tab:param_sim1}
\end{table}

For \emph{simulation 2} we choose interaction parameters 
$2\pi r_1^2t_1\log(\gamma_1)=5$  and $2\pi r_2^2t_2\log(\gamma_2)=-5$, 
as shown in the bottom left panel of Figure~\ref{fig:gamma_sim1-2}. 
The bottom right panel of Figure~\ref{fig:gamma_sim1-2} shows a
realization of the process with these parameters. We observe small scale 
clustering and large scale inhibition between points. The estimates of 
the parameters are given in Table \ref{tab:param_sim2}.

Note that Figure~\ref{fig:gamma_sim1-2} and 
Tables~\ref{tab:param_sim1}--\ref{tab:param_sim2} correspond 
to a single realization of the multi-scale area-interaction model
and one should be hesitant to draw any conclusions on the efficacy
or otherwise of the pseudo-likelihood method from this illustration.

\begin{table}[hbt]
\begin{center}
\begin{tabular}{|r|rrr|} 
\hline
                                 & Estimate   & 2.5 \% &  97.5 \% \\
\hline
$\log\lambda$	                 &  8.25  &  4.56 & 10.50 \\ 
$2\pi r_1^2t_1\log(\gamma_1)=5$   &  7.17  &  2.69 & 12.00 \\
$2\pi r_2^2t_2\log(\gamma_2)=-5$  & -2.39  & -6.40 & 1.19 \\
\hline
\end{tabular}
\end{center}
\caption{Parameter estimates for simulation 2.}
\label{tab:param_sim2}
\end{table}

\section{Data. Varicella in Valencia}
\label{section:data}

The \emph{Varicella-zoster virus} (VZV) is a highly contagious virus, spread 
worldwide, which causes two clinical syndromes: \emph{varicella}, also known 
as chickenpox, and \emph{herpes zoster}, otherwise known as shingles. In 
this paper we will focus on the spatial, temporal and spatio-temporal 
behavior of varicella. 

Varicella is transmitted from person to person by direct contact with the 
rash or inhalation of aerosolized droplets from respiratory tract secretions 
of patients with varicella. In temperate countries more than $90\%$ of the 
infections occur before adolescence and less than $5\%$ of adults remain 
susceptible. Varicella is mostly a mild disorder in childhood, but tends 
to be more severe in adults. The first symptoms of varicella generally 
appear after a $10$ to $21$ days incubation period. It is characterized by an 
itchy, vesicular rash, fever and malaise. Varicella is generally 
self-limited and vesicles gradually develop crusts. It usually takes about 
$7$ to $10$ days for all the vesicles to dry out and for the crusts to 
disappear. This gives us a time period, from infection to completely 
dried vesicles, between $17$ and $31$ days.  

Reported infection after household exposure ranges from $61\%$ to $100\%$ 
\cite{Gershon_08,WHO_book} which indicates small range interaction. The 
disease may be fatal, especially in neonates and immuno-compromised 
individuals. The epidemiology of the disease is different in temperate 
and tropical climates. The reasons behind this behavior may be related 
to climate, population density and risk of exposure \cite{conselleria,WHO}. 

In this paper we analyze varicella cases registered in Valencia, Spain, 
during 2013. Valencia is the third largest city in Spain with a population 
of around $800,000$ inhabitants in the administrative center ($19$ districts) 
and an area of approximately $134$ km$^2$ \cite{cityhall}. The study area 
is represented by districts $1$ to $16$. The remaining districts 
are very sparsely populated and are located far from the urban core. During 
the year 2013, $921$ cases of varicella were registered in the study area 
in the course of $52$ weeks \cite{conselleria}. 

\begin{figure}[hbt]
\centering\includegraphics[width=0.45\textwidth]{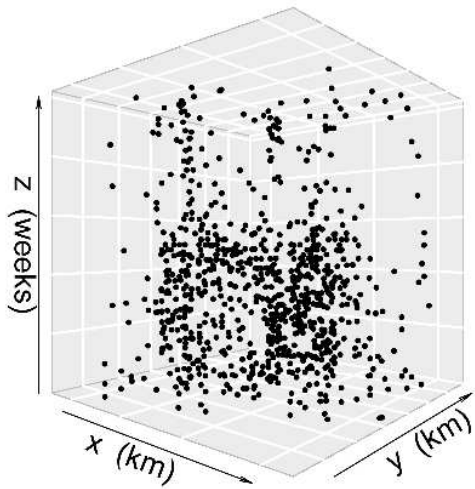}
\caption{Spatio-temporal pattern of weekly varicella cases in Valencia
during 2013, where the spatio-temporal study area is 
$W_S \times W_T = ([0,9] \times [0,9]) \times [0,52]$ (km$^2\times$ weeks).}
\label{fig:sp_t_varicela}
\end{figure}

The spatial coordinates of the varicella cases are expressed in latitude 
and longitude. First we transform them from longitude/latitude to UTM scale 
expressed in meters \cite{Snyder}. We then re-scale the spatial coordinates 
to kilometers such that the spatial study area reduces to $[0,9]\times[0,9]$. 
The temporal component of the process takes values from $0$ to $51$. For 
computational purposes to be explained later, we take the interval $[0,52]$ 
as the time window. Therefore, we set the spatio-temporal study area to 
$W_S \times W_T = ([0,9] \times [0,9]) \times[0,52]$ (km$^2\times$ weeks). 
The spatio-temporal pattern of all varicella cases thus obtained is shown in 
Figure~\ref{fig:sp_t_varicela}. The $x$- and $y$-axis represent the spatial 
coordinates in kilometers and the $z$-axis represents the time component in 
weeks. 

The main focus of our varicella data analysis is to quantify the 
interactions across a range of spatio-temporal scales. We do so by
using the spatio-temporal multi-scale area-interaction model introduced 
in Section~\ref{section:space-time area int}.

First we need to get some idea about a plausible upper bound to the values 
of the irregular parameters $(r_j, t_j)$, $j = 1, \ldots, m$, in 
model \eqref{model_with_r_and_t}.  To this end, we use summary statistics 
for the spatial and temporal projections of the space-time point pattern 
shown in Figure~\ref{fig:sp_t_varicela}.

\begin{figure}[hbt]
\centering
\includegraphics[width=0.45\textwidth, height=0.45\textwidth]{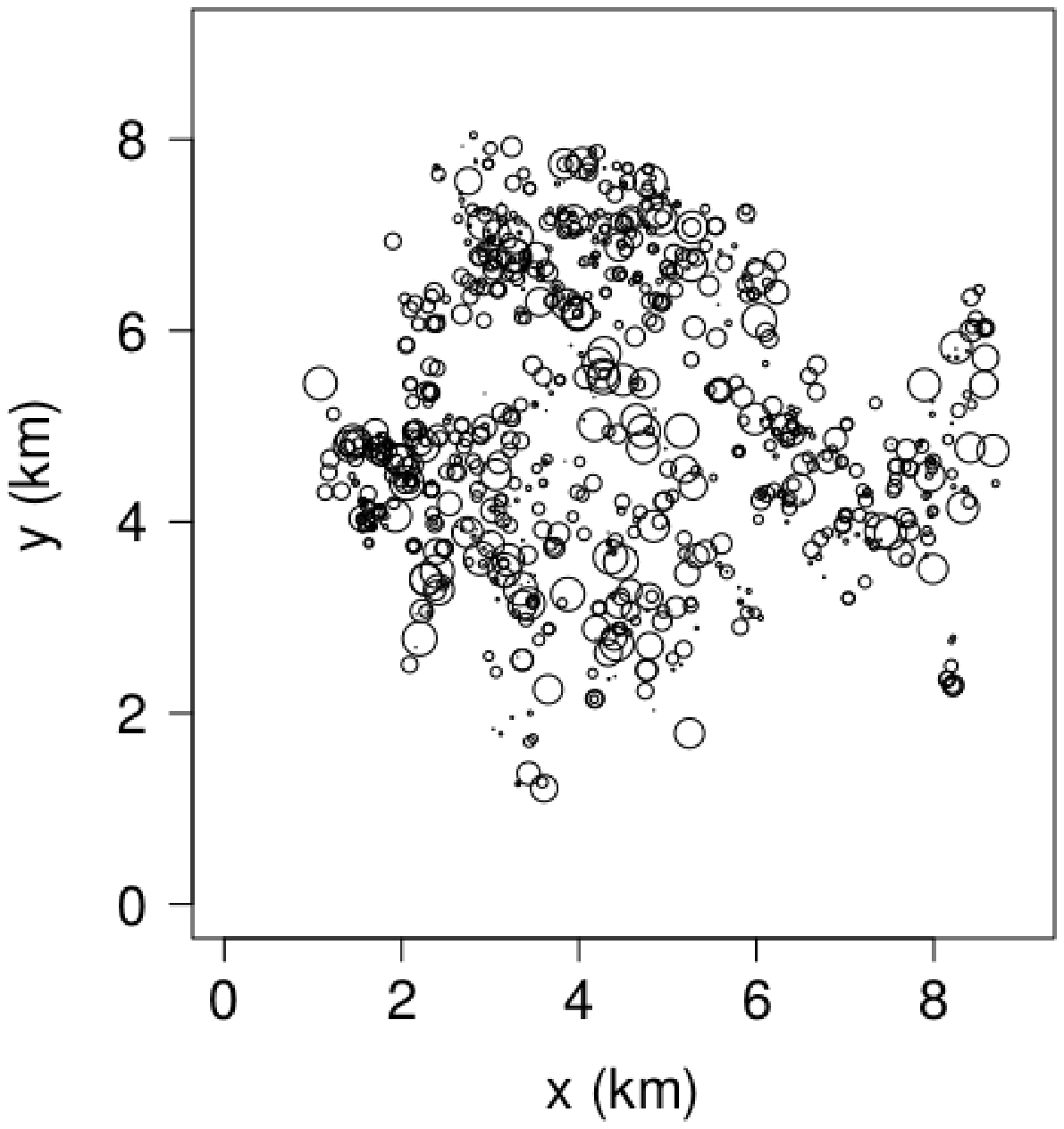}
\includegraphics[width=0.5\textwidth, height=0.45\textwidth]{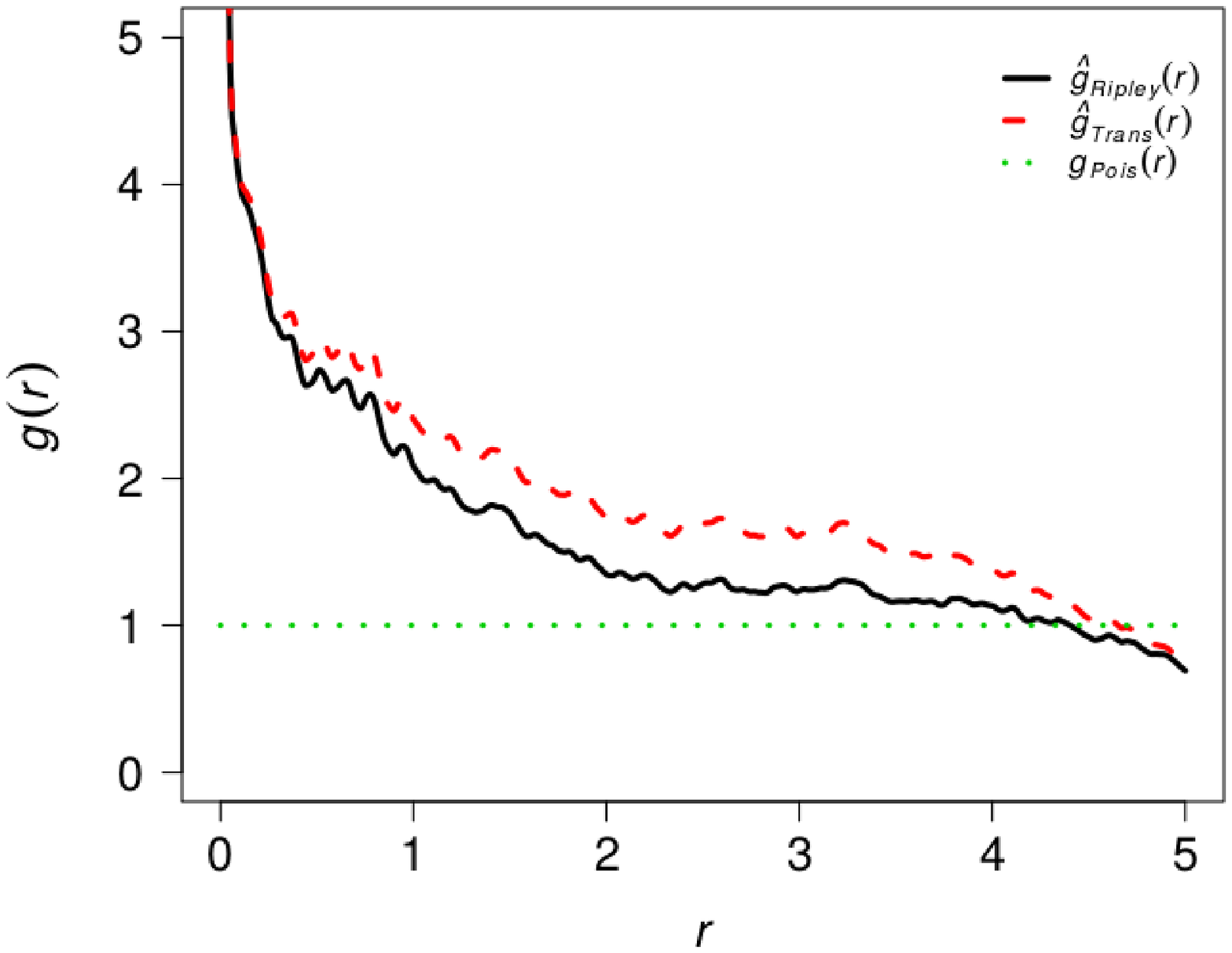}
\caption{(Left) Spatial projection of the spatio-temporal point pattern 
for the varicella data. After projection, locations were jittered 
using a maximum jitter distance of $20$ metres. 
(Right) Estimated pair correlation function for the jittered spatial 
point pattern shown in the left panel.}
\label{fig:sp_proj}
\end{figure}

The left panel in Figure~\ref{fig:sp_proj} shows the projection of all 
points onto the spatial region. The sizes of the circles are proportional 
to time, the bigger the circle, the more recent the event. Due to the 
projection, duplicate locations are observed, so we jitter the 
coordinates uniformly on the spatial region around the duplicated 
points using a maximum jittering distance of $20$ meters. 
To get a rough indication of the spatial interaction range, we pretend
that the pattern is stationary and isotropic, and estimate the pair 
correlation function. The result is shown in the right panel of
Figure~\ref{fig:sp_proj}.
Recall that for a Poisson process the pair correlation function is equal 
to $1$. Values of the pair correlation function lower than $1$ indicate 
inhibition and values larger than $1$ suggest clustering. 
Figure~\ref{fig:sp_proj} suggests that the pair correlation function is 
approximately constant from $2$ kilometers onward, which indicates a maximum
value for the $r_i$ of around $1$ kilometer. On a cautionary
note, we need to keep in mind that the estimator only takes into account 
the spatial pattern of points and assumes isotropy.

\begin{figure}[hbt]
\centering
\includegraphics[width=0.45\textwidth, height=0.40\textwidth]{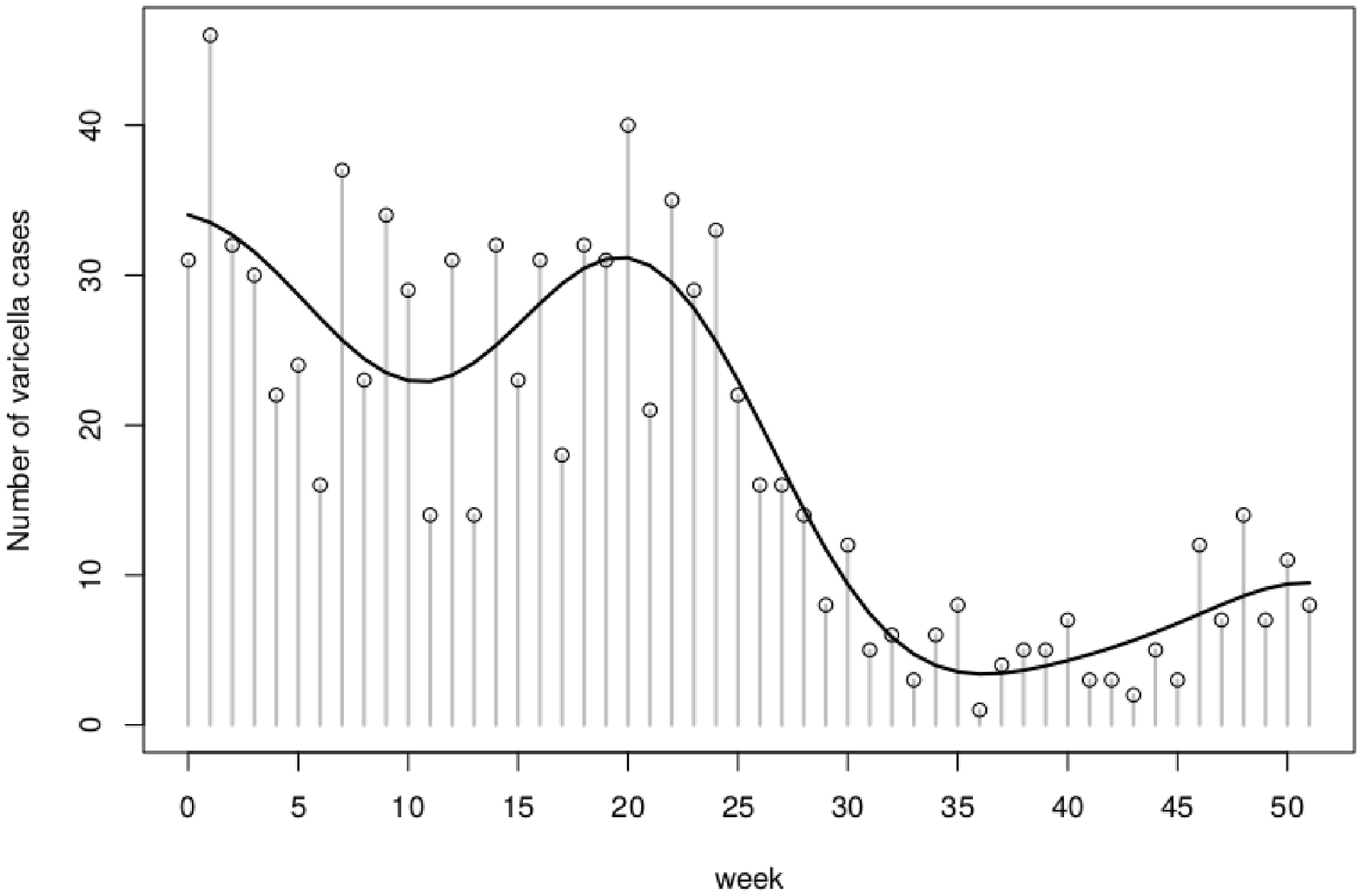}
\includegraphics[width=0.40\textwidth, height=0.40\textwidth]{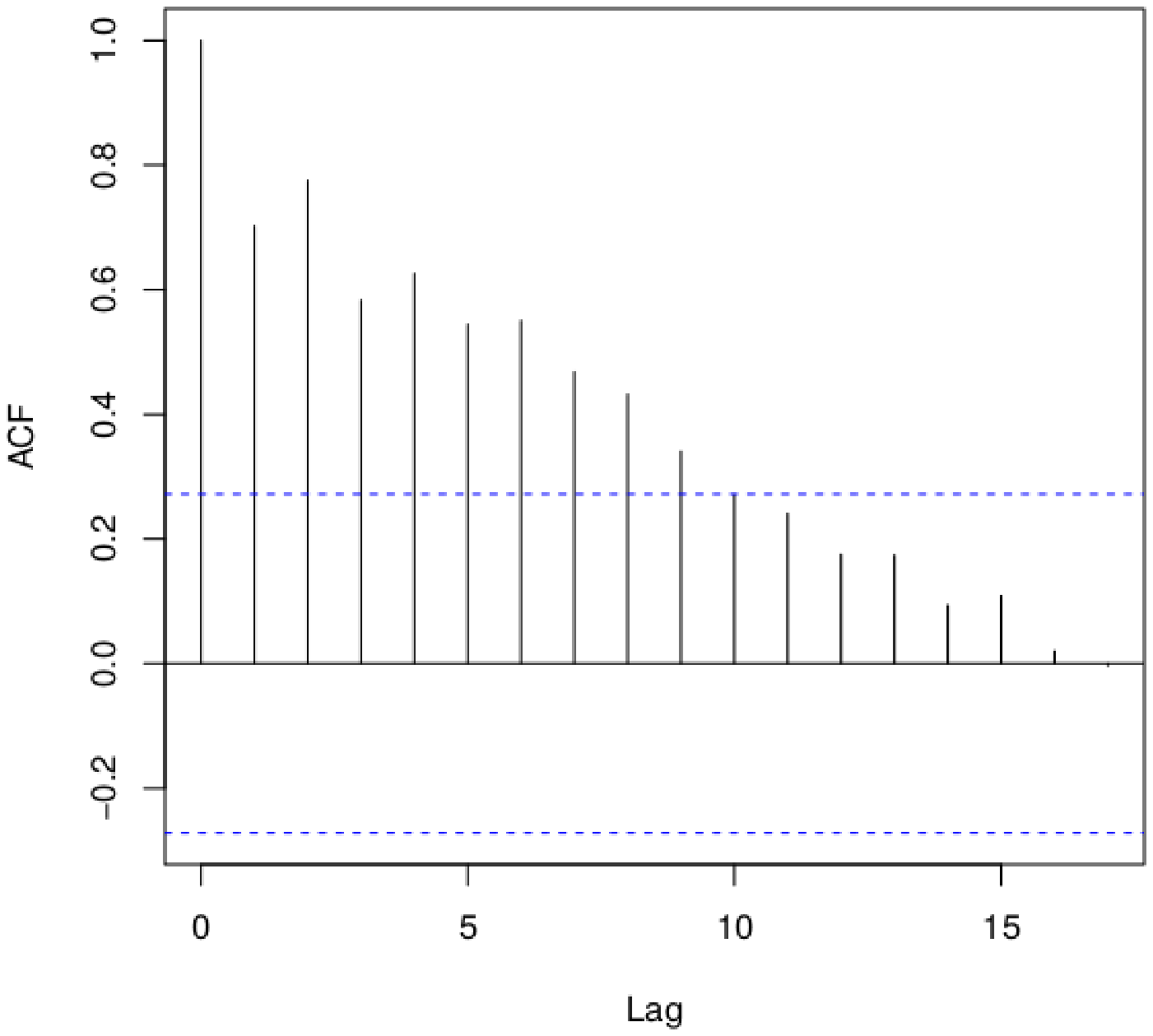}
\caption{(Left) Weekly reports of varicella cases ($\circ$) and fitted 
regression curve (--). 
(Right) Estimated auto-correlation function for the data shown in 
the left panel.}
\label{fig:time_pattern}
\end{figure}

The left panel in Figure~\ref{fig:time_pattern} shows the temporal 
evolution of varicella over the $52$ weeks, where the small circles 
$\circ$ represents the number of registered cases. The right panel
displays the estimated auto-correlation function which measures the 
correlation between the values of the series at different times as a 
function of the time lag between them. Figure~\ref{fig:time_pattern} 
suggests possible correlation for time lags as big as $15$ weeks. 
This gives us an estimate for the maximum value for the $t_i$ of about 
$7.5$ weeks. Note that caveats similar to the spatial case apply.

Now that we have estimated the maximum spatial and temporal range for the 
model, the following step in our analysis is to consider covariate information.
The most important factor in the transmission of any kind of disease, and 
especially a highly contagious one such as varicella, is the population. In 
areas with very low population we will probably not register as many varicella 
cases as in highly populated areas. Thus, the pattern of varicella cases can 
drastically change from one area to another, depending on the spatial 
distribution of the population, and from one week to the next one. We 
express the spatio-temporal inhomogeneity term in 
equation~\eqref{model_with_r_and_t} as a product
$\lambda(x,t) = \lambda(x) \, Z(t)$, $x \in [0,9]^2$, 
$t \in \{ 0, \ldots, 51 \}$, between a non-parametric estimate of the 
population density $\lambda(x)$ and a re-scaled parametric estimate of 
the temporal component $Z(t)$.

First consider the spatial component $\lambda(x)$. The population data 
available to us consist of the number of people living in each census 
section of the city of Valencia, a total number of $559$ sections 
(districts $1$ to $16$). We randomly generate within each section $p$ 
points, where $p$ is equal to the number of people living in that 
particular section. This way, we obtain a sample of the population for 
the city of Valencia. We estimate its intensity by a kernel estimator, 
keeping in mind that the bandwidth has to be chosen carefully, to
get $\lambda(x)$, $x\in W_S$.
 
Following \cite{iftimi2014} we fit a harmonic regression 
to the pattern of the weekly varicella counts
\begin{align}\label{eq:temporal_regression}
Z(t) = c_0 +  \sum_{j=1}^3 \left(
c_j \cos( 2 \pi j t / 52 ) + d_j \sin( 2 \pi j t / 52 )
\right) + c ( a + b t ),
\end{align}
where $Z(t)$ denotes the number of varicella cases at time $t$, 
$t = 0, \ldots, 51$, and $c_0, a, b, c$, $c_j, d_j$, $j = 1, 2, 3$, 
are the parameters of the model.

The left panel in Figure~\ref{fig:time_pattern} shows the fitted 
regression curve. We observe a period at the beginning of the year, 
from winter until spring, with large numbers of varicella cases, 
and a second period starting around week $26$, in which the number
of cases decreases. These periods correspond roughly with the school 
term and the summer break. Also, in 2013, in Spain, 
there were several holidays besides the summer and winter holidays. 
On March 19, San Jose is celebrated and the period from the 24th to 
the 31st of March corresponds to the Easter holidays. As a consequence 
we can observe in Figure~\ref{fig:time_pattern} a decrease during the 
11th and 12th week. Towards the end of the year, the number of cases
picks up again as the Michaelmas term begins.

Finally, we re-scale the parametric estimate of the temporal component
$Z(t)$ by $100$, in order to avoid obtaining extreme values for the
spatio-temporal inhomogeneity term $\lambda(x,t)$.

Since realizations of \eqref{model_with_r_and_t} do not contain 
points with equal time stamps, we jitter in time as well as space.
More precisely, the week index is replaced by a time stamp that is
uniformly distributed in the indicated week so that the temporal 
component falls in $W_T = [0,52]$. To estimate the parameters, we 
follow the steps described in Algorithm~\ref{algorithm_mpl}. For
constructing the quadrature points we partition $W_S\times W_T$ 
into $9\times 9\times 52=4,212$ cubes of equal volume $1$ and place 
one dummy point in the center of each cube. Doing so, we obtain a 
total of $5,133$ dummy and data points. 
We attribute to each point a weight equal to the volume of the cube divided
by the number of dummy and data points inside the cube containing the point. 
We then compute the sufficient statistics $S_j(\cdot,\cdot)$ corresponding 
to each point using the \texttt{MPPLIB C++} library of \cite{mppblib}. We 
follow Algorithm \ref{algorithm_mpl} and obtain estimates for the parameters 
$\gamma$. The analysis and visual representations have been carried out using 
the statistical software \texttt{R} \cite{R_software} together with the 
\texttt{spatstat} \cite{baddeley_book}, \texttt{plot3D} \cite{plot3d} and 
\texttt{rgdal} \cite{rgdal} packages.

Recall that we found indications for the maximum spatial range to be
about $2$ kilometers, the maximum temporal range $15$ weeks. As suggested in 
\cite{baddeley_turner} we fitted the model for a range of values
$(r_j,t_j)$, $j=1, \dots, m$, in the larger domain $[0,2] \times [0,15]$
and for different $m\in\{3,4,5,6,7,8\}$ to choose the optimal combination. 

\begin{figure}[hbt]
\centering
\includegraphics[width=0.45\textwidth]{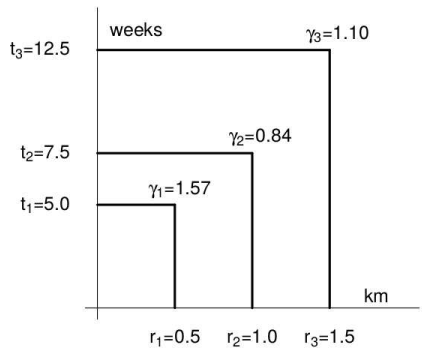}       
\caption{Model parameters for the varicella data.} 
\label{fig:multi_scale_gamma_data}
\end{figure}

We estimate $m=3$, that is, three spatio-temporal scales and the 
corresponding parameters. For the spatial scales we selected $r_1 = 0.5, 
r_2 = 1$ and $r_3 = 1.5$ kilometers and for the temporal scales $t_1 = 5$, 
$t_2 = 7.5$ and $t_3 = 12.5$ weeks. Figure~\ref{fig:multi_scale_gamma_data} 
shows the multi-scale interaction in the data together with the estimated 
values of the model parameters. Also, Table~\ref{tab:param_data} shows the 
estimated parameters of the model together with a confidence interval.

\begin{table}[hbt]
\begin{tabular}{|rrrrrr|}
\hline
    &   Spatial scale & Temporal scale & Parameters & 2.5\% & 97.5\% \\
\hline
(Intercept) &  &  & 1.20 & 1.09 & 1.31  \\
            &   0.5   & 5.0   & 1.57 & 1.39 & 1.78  \\
            &   1.0   & 7.5   & 0.84 & 0.74 & 0.95  \\
            &   1.5   & 12.5  & 1.10 & 1.00 & 1.23  \\
\hline
\end{tabular}
\caption{Parameter estimates for the varicella data.}
\label{tab:param_data}
\end{table}

As stated before, the time period from infection to completely dried 
vesicles is between approximately $17$ and $31$ days. In 
the fitted model we observe that for a spatial lag of $0.5$ kilometer and 
a temporal lag of $5$ weeks there is clustering (significant 
$\gamma_1 = 1.57$). This means that for a period of five weeks and at 
rather small distance (as far as $0.5$ kilometers), a phenomenon of 
aggregation is observed between cases of varicella. The time lag 
corresponds more or less with the period of $31$ days indicated by 
the epidemiologists. This is caused by the main feature of chickenpox, 
being a contagious disease. 

The fitted model also exhibits inhibition for spatial lags as far as $1$ 
kilometer and temporal lags up to $7.5$ weeks (significant $\gamma_2 = 0.84$).
This might be a result of the fact that after recovery from varicella,
patients usually have lifetime immunity. For higher spatial and 
temporal lags the model suggests no interaction ($\gamma_3 \approx 1$), which 
corresponds with the rough bounds we found before and is in accordance with 
the beliefs of the epidemiologists. If you are situated 
far away from a varicella case, both in space and in time, you are less 
susceptible to contract the disease due to the contagious factor. Also, 
the probability of contracting the disease would be the same as the 
incidence of varicella.

\begin{figure}[hbt]
\centering
\includegraphics[width=0.45\textwidth]{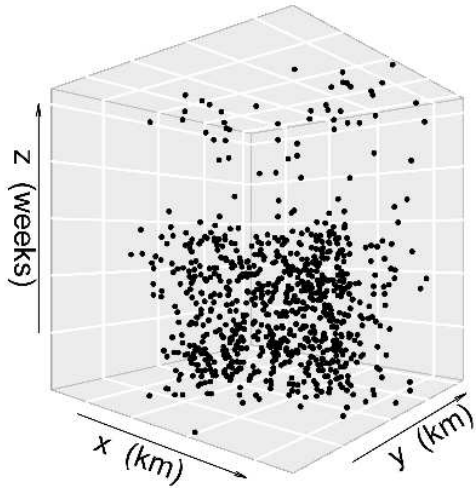}
\caption{Realization from the model fitted to the varicella data.}
\label{fig:sim_data}
\end{figure}

To validate our model, we simulated a number of space-time multi-scale 
area-interaction processes with the fitted parameters using the 
Metropolis-Hastings algorithm described in 
Section~\ref{subsection:mh_algorithm} for $20,000$ iterations, which
seems enough for the algorithm to converge based on diagnostic plots.
Figure~\ref{fig:sim_data} shows one such simulation. Comparing 
Figure~\ref{fig:sp_t_varicela} and \ref{fig:sim_data}, we note
that the simulated spatio-temporal point pattern is similar to the varicella 
point pattern.

\section{Discussion and final remarks}
\label{section:discussion}

In this paper we developed an extension of the area-interaction model 
that is able to incorporate different types of interaction at different 
spatio-temporal scales and proposed methods to simulate this process. 
We discussed inference and demonstrated the pseudo-likelihood method 
on simulated data. Additionally, we analyzed a spatio-temporal point 
pattern of varicella in the city of Valencia, Spain. For future work, 
it would be interesting to apply our model to other diseases that may 
exhibit interaction at several scales in space and time. It would also 
be very interesting to apply this model to data that are not necessarily 
related to epidemiology. Earthquake patterns, for instance, tend to show 
aggregation but also inhibition at different scales. Indeed, we believe 
that the proposed model may find applications in a wide range of research
fields, such as forestry, geology and sociology.

As stated in Section~\ref{section:data}, varicella is a highly infectious 
disease. We are certain that, in addition to the effect of the population, 
there are other covariates that may influence the spatio-temporal behavior 
of the disease. Therefore, an important goal for future work is to consider 
adding covariates that can improve the model. For example, \cite{WHO} suggests 
that there are some climatic factors that can influence the epidemiology of 
varicella. Thus, covariates such as the monthly average temperatures, weekly 
average levels of rainfall, average hours of sunshine, or other climate 
related covariates, may provide useful information to the analysis of 
varicella. Also, additional information on the composition of households, 
income per capita or other socio-economical covariates might improve the model. 
Another important covariate that could be taken into account in future work is 
related to the locations of kindergartens, schools and high-schools: The 
distance from a case to the nearest school may provide important information 
for the analysis of varicella.

\section{Software}
\label{software}

Software in the form of \texttt{R} code and complete documentation are
available on request from the corresponding author (iftimi@uv.es).

\section*{Acknowledgments}

The authors thank Francisco Gonz\'{a}lez of the Surveillance Service and 
Epidemiological Control, General Division of Epidemiology and Health 
Surveillance -- Department of Public Health, Generalitat Valenciana for 
providing the varicella data. They are grateful to Professor R.\
Turner for helpful discussions. 

The first author gratefully acknowledges financial support from the 
Ministry of Education, Culture, and Sports (Grants FPU12/04531 and 
EST15/00174) for a research visit to The Netherlands. She thanks the 
Stochastics group at CWI and the SOR-chair at Twente University 
for their hospitality. 

\bibliographystyle{plain}

\end{document}